\documentstyle[12pt,aaspp4]{article}

\def\ace{$\alpha_{\rm CE}$}
\def\snia{SN~Ia}
\def\logm{log~$M/M_\odot$}
\def\mch{$M_{\rm ch}$}
\def\msun{$M_\odot$}

\def\lta{\lower 0.5ex\hbox{$\buildrel < \over \sim\ $}}
\def\gta{\lower 0.5ex\hbox{$\buildrel > \over \sim\ $}}
\def\twid{$\sim$}
\def\halpha{H$\alpha$}
\def\kms{km~s$^{-1}$}
\newcommand\tento[1]{$10^{#1}$}
\newcommand\scinote[2]{#1~$\times$~10$^{#2}$}
\newcommand{\lam}[1]{$\lambda{#1}$}

\def\fd{\hbox{$.\!\!^{\rm d}$}}
\def\fh{\hbox{$.\!\!^{\rm h}$}}
\newcommand{\singlespacing}{\renewcommand{\baselinestretch}{1}}

\begin{document}

\title{Close Binary White Dwarf Systems: \\ Numerous New Detections
and Their Interpretation}
\author{Rex A. Saffer\altaffilmark{1} \& Mario Livio}
\affil{Space Telescope Science Institute\altaffilmark{2}, 3700 San
Martin Dr., Baltimore, MD 21218}
\affil{email: \em{saffer@stsci.edu, mlivio@stsci.edu}}
\author{Lev R. Yungelson}
\affil{Institute of Astronomy of the Russian Acadamy of Sciences, \\
48 Pyatnitskaya St., Moscow 109017, Russia}
\affil{email: \em{lry@inasan.rssi.ru}}
\bigskip
\affil{Accepted for publication in the Astrophysical Journal}

\altaffiltext{1}{Current address: Department of Astronomy \& Astrophysics,
Villanova University, 800 Lancaster Ave., Villanova, PA 19085,
email: \em{saffer@ast.vill.edu}}

\altaffiltext{2}{Operated by The Association of Universities for
Research in Astronomy, Inc., for the National Aeronautics and Space
Administration.} 

\begin{abstract}
We describe radial velocity observations of a large sample of
apparently single white dwarfs (WDs), obtained in a long-term effort
to discover close, double-degenerate (DD) pairs which might comprise
viable Type Ia Supernova (\snia) progenitors. We augment the WD sample
with a previously observed sample of apparently single subdwarf B
(sdB) stars, which are believed to evolve directly to the WD cooling
sequence after the cessation of core helium burning. We have
identified 18 new radial velocity variables, including five confirmed
sdB+WD short-period pairs. Our observations are in general agreement
with the predictions of the theory of binary star evolution. We
describe a numerical method to evaluate the detection efficiency of
the survey and estimate the number of binary systems not detected due
to the effects of varying orbital inclination, orbital phase at the
epoch of the first observation, and the actual temporal sampling of
each object in the sample. Follow-up observations are in progress to
solve for the orbital parameters of the candidate velocity variables.
\end{abstract}

\section{Introduction}

Both theoretical and observational investigations into the origins of
Type Ia Supernovae (\snia) have focused on binary star
evolution. Within this framework, scenarios are divided into ``single
degenerate'' (\cite{wheliben}) and ``double degenerate''
(\cite{rwebb}; \cite{ibentut84a}) models.  In the former models, a
white dwarf (WD) accretes from a nondegenerate companion, while in the
latter, two WDs with a total mass exceeding the Chandrasekhar limit
(\mch) are assumed to coalesce.  The latter scenario predicts that
there should exist a population of close, short-period ($P$~\lta
10$^{\rm h}$) binary WD systems, or double degenerates (DDs), with
total system masses in excess of \mch.  The binary components are
initially brought to short-period orbits by common envelope (CE)
evolution, with the energy necessary to expel the envelope taken from
orbital shrinkage (\cite{ibenlivio}). Subsequently, further loss of
orbital angular momentum through emission of gravitational wave
radiation provides the mechanism through which the merger can take
place. It is important to note that at present it is not entirely
clear if the coalescence of two WDs with a total mass exceeding \mch\
really produces a \snia\ or rather an accretion induced collapse
(cf. \cite{mochkov}; \cite{mochetal}). However, the existence of DD
systems is a definite prediction of the theory of close binary star
evolution.

Recently, several authors have made the suggestion that the properties
of observed \snia\ are not consistent with the detonation of
progenitors having a single, well-defined mass. These include an
apparently real dispersion in the peak \snia\ absolute B magnitudes
and a correlation between the peak magnitude and the decay rate of the
B light curve (\cite{phillips}; \cite{hamuy}). \cite{pinto} report
matching the relationship only by requiring a variation in the amount
of carbon burned to Ni$^{56}$ of nearly a factor of two. It has been
suggested that accumulation of He atop an ordinary sub-\mch\ CO WD, or
the merger of sub-\mch\ He+CO WDs might provide the necessary range of
masses and produce edge-lit detonations (ELDs) with \snia\
characteristics (e.g., \cite{livne}; \cite{woosley};
\cite{jwletal96}). The viability of this mechanism is controversial
(see, e.g., \cite{branch}; \cite{wheel}), and existing estimates of
the expected rate of ELDs in sub-\mch\ systems fall short of the
inferred Galactic \snia\ rate (\cite{tutyung96}). Another alternative
is that some of these systems may evolve into ultra-short period
interacting AM CVn systems, which are believed by many authors to
consist of lower-mass CO white dwarfs accreting from very low-mass He
WD secondaries (e.g., Liebert et al. 1996).

\cite{slo} discovered the prototype DD, L~870--2 (EG 11), which has an
orbital period $P \sim 1\fd 56$.  However, the period is too long for
the system to merge on an astrophysically interesting time scale. When
the merger does take place, a \snia\ probably will not result since
the total system mass is below \mch\ (\cite{eg11}). In recent years,
three systematic searches for DDs have been performed. Two of these
(\cite{robshaft}, hereafter RS; \cite{fossetal}, hereafter FWG) failed
to discover any candidates at all. The third (\cite{bragetal},
hereafter BGRD) did discover one DD (and 4 other candidates), but the
low primary mass ($M$ \twid 0.37~\msun; \cite{brag95}) of the
confirmed DD renders it an unlikely \snia\ progenitor. We note that
BGRD found a period $P$ \twid~1\fd 18 for their confirmed DD
(0957$-$666), however, recent observations (Moran, Marsh, \& Bragaglia
1997) reveal that the system actually has a period $P$ \twid 88
minutes and primary and secondary masses of 0.37 and 0.32 \msun,
respectively. Recently, observations targeting the low-mass tail
of the WD mass distribution of Bergeron, Saffer, \& Liebert (1992)
spectacularly confirmed the prediction that WDs with masses less than
the canonical core-helium ignition mass are likely to be post-common
envelope objects residing in short-period binary systems (Marsh,
Dhillon, \& Duck 1995; Marsh 1995). However, the low WD masses again
make them unlikely \snia\ progenitors. These negative results have
cast some doubt on the merging DD scenario as the favored mechanism
for producing \snia.

There does exist one example of a DD system where the combined mass is
believed to exceed the Chandrasekhar limit. \cite{jwletal92}
discovered that LB~11146 consists of a normal DA WD and a highly
magnetic DAXP WD. The combined mass of the system is between 1.7--1.9
\msun, but the orbital period is unknown. \cite{glennetal} failed to
detect radial velocity variations in the core of \halpha\ to a limit
\twid 20 \kms\ with periods \twid 1 hour, and a recent {\em Hubble
Space Telescope} imaging program (\cite{gds98}) did not resolve the
components to a separation $\leq$ 0\farcs 025, corresponding to $\leq$
1 a.u. on the plane of the sky. If the progenitor stars were massive,
as seems likely given the large masses of the remnants, close binary
interaction is likely to have occurred in a previous AGB
phase. Further work to establish the orbital period of this remarkable
binary is urgently needed.

Another possible outcome of close binary evolution is a merger during
the late stages of a CE phase leading to the formation of a single,
possibly massive white dwarf (e.g., \cite{lps92};
\cite{segretal}). Large numbers of hot white dwarfs were detected in
the ROSAT and EUVE surveys, and analyses of optical follow-up
observations (e.g., \cite{mmarsh}; \cite{vennes}; \cite{finley})
determined stellar parameters, including masses. Comparing the ROSAT
and EUVE samples with the cooler, optically selected sample of
\cite{bsl} revealed a statistically significant excess of hot,
supermassive WDs, which may represent a population of coalesced binary
WDs. The high-mass excess may partly arise from differences in the
cooling rates of normal (\twid 0.6 \msun) and massive ($>$ 1.0 \msun)
stars, but there may still be too many massive white dwarfs to be
explained by the predictions of single star evolution, with a
conventional initial mass--final mass relation, initial mass function,
and star formation history.

Recent theoretical work (\cite{yungetal}, hereafter YLTS;
\cite{ibtutyung}, hereafter ITY) reveals that consideration of the
"visibility function", a consequence of WD cooling along evolutionary
sequences, is an important factor in calculations of the expected DD
number distribution. Systems born with orbital periods $P$~\gta
10$^{\rm h}$ do not evolve to shorter periods by emission of
gravitational radiation on interesting (i.e., short) timescales. Below
10$^{\rm h}$, an approximate equilibrium is established between
systems born into short-period orbits and those that vanish either by
merging or by cooling in some \tento{8} years beyond the detection
limits of surveys with relatively bright limiting magnitudes, such as
our own. Only a small fraction of these pairs will actually merge in
the \tento{8} year period of visibility, and only \twid 1/20 of the
pairs with $M_{\rm tot}$ \gta \mch\ are visible (cf. ITY, Figure
2). These results indicate that previous DD searches had sample sizes
too small to reveal the massive, short-period systems of interest.  On
the other hand, it is clear that the close DDs that {\em were\/}
discovered (Saffer et al. 1988; BGRD; \cite{tmarsh1}; \cite{tmarsh2})
fall within the peak of the expected DD number distribution (cf. YLTS,
ITY). YLTS suggested that further searches with increased sample sizes
are likely to reveal the existence of the short-period, massive DD
population. In this paper, we report observations of an enlarged
sample in order to place meaningful constraints both on DDs as \snia\
progenitors and, more generally, on the theory of close binary star
evolution.

\section{Observations and Radial Velocity Variability}

Our sample comprises 107 apparently single DA (hydrogen line) and DB
(helium line) WDs with $V \leq 15$ drawn from the catalog of
\cite{mccook}, which are largely not in common with the previous three
systematic surveys. Over two observing seasons, in a total of 21
nights in 4 observing runs from 1994--1996, we have obtained radial
velocity spectroscopy at the Kitt Peak National Observatory's 2.1-m
reflector equipped with GOLDCAM. This instrumentation provided 3\AA\
FWHM spectral resolution over a spectral range of \twid 700\AA\
centered on \halpha\ and including He {\sc i} \lam{6678}. Exposure
times were chosen to achieve a signal-to-noise (S/N) ratio of \twid 30
per pixel.

We have augmented the WD sample with a sample of 46 apparently single
subdwarf B (sdB) stars previously observed by Saffer (1991; see also
\cite{sbkl}). The radial velocity observations were obtained using the
Steward Observatory's Kitt Peak Station 2.3-m reflector, equipped with
instrumentation similar to that used at the Kitt Peak 2.1-m and
providing nearly identical spectral coverage and resolution. The
inclusion of the sdB stars in the sample is appropriate since these
core helium-burning stars are believed to evolve directly to the WD
cooling sequence on a relatively short time scale following core
helium exhaustion. Theoretical calculations which predict the period
and mass distributions of DDs do take into account the contribution
from the descendents of He stars.

Our choices of spectral features, resolution, and S/N ratio for our
observations convey several advantages: 1) In all but the hottest DA
WDs, and in all sdB stars, \halpha\ has a sharp (FWHM \twid 1\AA)
non-LTE core, making it possible at our resolution and S/N ratio to
measure radial velocities with 20--30 \kms\ errors despite the extreme
width of the Stark-broadened absorption line. The He {\sc i}
\lam{6678} line in DB WDs provides similar precision, 30--40 \kms. The
close, short-period DDs of interest have peak-to-peak radial velocity
amplitudes nearly an order of magnitude larger than these detection
limits, and any existing systems would have a significant probability
of detection (for our sample, \twid 70\% in the period range 2$^{\rm
h}$--20$^{\rm d}$; see \S 7 for details). 2) Our method detects
objects with similar temperatures, like the prototype L~870--2, by
revealing variations of both absorption cores in a double-lined
spectrum. The technique of differential photometry in the wings of
the absorption line does not detect this type of system. 3)
Spectroscopic observations in the core of \halpha\ offer the maximum
velocity resolution for any WD spectral features observable from the
ground at intermediate resolution. 4) The large throughput at
intermediate spectral resolution makes it possible to observe large,
relatively faint samples with modest telescope apertures.

Our targets were observed twice on any given night separated by 1--2
hr, followed by a third observation a day or two later. Radial
velocity variations of short-period systems are easily detected simply
by comparing spectra by eye, although in follow-up observations to
determine the orbital parameters, velocities will be measured
formally by line profile fitting. In particular, derived periods and
velocity semi-amplitudes will determine mass functions and provide
lower bounds on total system masses.

In Table \ref{wdsample}, we list all the WDs in our sample. Successive
columns give the WD number, name, $V$ magnitude, ($B$--$V$) and
($U$--$B$) colors when available in the literature (occasionally only
Greenstein multichannel or Str\" omgren colors are available), and
references. We also list 7 objects (Note \#3) from samples of sdB
stars (Saffer 1991; \cite{sbkl}), which show significant velocity
variability. The remaining non-variable sdB stars and their weighted
mean velocities and errors are given in Table \ref{sdbsample}. In
Table \ref{ddvariables} we list candidate DD systems which show
velocity variability identified by visual examination of stacked
spectra, the spectroscopic analog of blinking photographic plates.
The last column indicates a weight of either 1 or 2, indicating
definite or marginal velocity variability, respectively. We identify
definite variables as those for which the spectral line cores vary by
\gta 1.5\AA, a 2.5$\sigma$ detection, on average. Marginal variables
did appear to vary, but at a lower level. The references are the same
as in Table \ref{wdsample}.

In the first 13 panels of Figure \ref{multistack}, we show spectra of
our best candidates (weight 1), emphasizing the cores of the \halpha\
lines. The sdB velocity variables are excluded from the figure --
unfortunately, the spectra could not be recovered from an ancient
archive tape. However, variability in the sdB stars was originally
measured by formal line profile fitting (see Table \ref{sdbsample} for
the results). The last two panels at the bottom right in Figure
\ref{multistack} show the detection of narrow, variable \halpha\
emission in the cores of the WD absorption lines, betraying 
otherwise hidden low-mass main sequence (MS) companions in
detached, pre-cataclysmic binary (PCB) systems. The \halpha\ emission
likely results from reprocessing of UV photons from the hotter WD in
the cooler stars' atmospheres (cf. \cite{ras93}).
 
The laboratory wavelength of \halpha\ is indicated with a vertical
line in each panel. The figure illustrates the ease with which
velocity variables may be identified, even in intermediate-dispersion
spectra, and despite the highly Stark-broadened \halpha\ absorption
wings.  We have detected a total of 23 definite variables, including 7
sdB+WD pairs and 2 white dwarf plus main sequence (WD+MS) pairs. Of
the total, 11 WD+WD pairs, all 7 sdB+WD pairs, and 1 WD+MS pair are new
detections, most of which require follow-up observations to solve for
the orbital parameters. Some follow-up observations have already been
accomplished -- the previously detected sdB velocity variables Feige
36 and Ton 245, both pre-white dwarfs, have been shown to be close
binary systems with periods of 8\fh 5 and 2\fd 5, respectively (Saffer
et al. 1998, in preparation). In addition, the sdB stars 0101+039,
1432+159, and 2345+318 have recently been shown to have a periods of
16\fh 1 (with a possible alias of 13\fh 7), 5\fh 39, and 5\fh 78,
respectively (Moran et al. 1998, in preparation). Low-mass MS
companions are ruled out by large mass functions and the non-detection
of infrared excesses. The companions are thus inferred to be compact,
most likely white dwarfs.

\section{Objects in Common with Previous Surveys}

Approximately 35\% of the objects in our survey are in common with
previous surveys. However, most of these objects have not been sampled
in the period range in which our survey has the greatest sensitivity:
\begin{enumerate}
\item 
The largest area of overlap with previous surveys is with RS (who
found no variables), with 26 of the objects listed in Table
\ref{wdsample} being in common. However, the RS temporal sampling
provided sensitivity to orbital periods between 30$^{\rm s}$ and
3$^{\rm h}$, while our survey is most sensitive to orbital periods
between 2$^{\rm h}$ and 10--20$^{\rm d}$. There is thus no substantive
overlap between the two samples in their common orbital period
ranges. Table \ref{ddvariables} shows that we detected 7 of these
objects as radial velocity variables, 4 marginally and 3 definitely.
\item 
There are 12 objects in Table \ref{wdsample} which overlap with FWG,
who also found no variables. Their temporal sampling provided
sensitivity to orbital periods between 3--10$^{\rm h}$.  Of the 12 FWG
objects, 6 are also in common with RS. Of the remaining 6 objects, we
have detected 2 as velocity variables, one marginally and one
definitely.
\item The overlap with the survey of BGRD, much of which took
place in the southern hemisphere, comprises only 10 objects, 4 of
which are also in common with RS and FWG. Of the remaining 6 objects,
we definitely detect 2 as velocity variables. These two objects were
not detected as variables by BGRD. 
\end{enumerate}

\section{Notes on Individual Objects}

\noindent {\em 0553+053.} This object is a strongly magnetic DA WD,
whose \halpha\ absorption line is widely split by the Zeeman effect
(\cite{gdsps}). We detect possible line profile variations in our
spectra which might be indicative of rotational or orbital
modulation. The velocity is otherwise very difficult to measure due to
the extremely broad and shallow absorption components.

\noindent {\em 0710$+$741, 1101$+$364, 1317$+$453, 1713$+$332,
2032$+$188.} These objects were detected as short-period binaries by
Marsh, Dhillon, \& Duck (1995) and Marsh (1995), and as such do not
represent new detections. Our survey was performed between November
1994 and April 1996, during which time we targeted these objects prior
to having knowledge of their duplicity. We detected variability in all
but one relatively faint object (2032+188; $V$ = 15.34), for which our
spectra had very poor S/N in the maximum exposure time of 30
minutes. This limit was imposed to avoid smearing spectra of very
short-period systems. Our observations of these objects are very
useful, in that they indicate that our detection of known variables is
very efficient, at least 80\% in this case. Also, since they were
observed independently they may be retained in the statistical
analysis.

\noindent {\em 0713+584, 1756+827.} These two objects have large
blueshifts, \twid 300 \kms, and are likely members of the Galactic
halo.

\noindent {\em 0900+554, 0940+068, 1247+553, 2204+071.} The first
three objects are classified as DA WDs in the catalog of
\cite{mccook}, but our observations clearly identify them as sdB or
blue horizontal branch (BHB) stars, as seen in the top three spectra
in Figure \ref{wd2sdb}. For comparison, in the bottom spectrum we show
a normal DA WD. The lower-gravity stars have narrow, sharp \halpha\
absorption lines, and He {\sc i} $\lambda$6678 is also detected. The
strength of the latter feature in 0900+554 and 1247+553 implies that
these two objects could be slightly cooler, lower-gravity BHB stars
and not sdB stars, as the latter in general tend to have very weak He
{\sc i} $\lambda$6678 absorption (cf. \cite{sbkl}). This is
certain for 0900+554, for which an existing blue optical spectrum
shows Balmer absorption extending to $n$ \gta 14. However, all three
objects will eventually evolve to low-mass (\twid 0.5 \msun) WDs and
rightfully remain included in the sample. The PG coordinates for
0900+554 (RA 09:00:13.4, DEC +55:26:42) were found to be in error --
the object's location is RA 09:00:05, DEC +55:28:50 (1950). 1247+533
is a visual double -- the sdB/BHB star is the southeast
object. 0940+068 is barely resolved as a visual binary in the
telescope guide star camera. It is not identified as composite in the
PG survey and appears single on the PG finder plate. The eastern
object is an sdB star, the western object a late-type main sequence
star. The fourth object, 2204+071, is ambiguously identified in the PG
survey, as there are actually two objects marked on the PG finder
plate. The southern object is a DA white dwarf analyzed by
\cite{bsl}. We serendipitously observed the northern object thinking
it to be the DA. Instead, we obtained the fourth spectrum from the top
in Figure \ref{wd2sdb}. The object is either a BHB star, in which case
the extreme strength of He {\sc i} $\lambda$6678 would imply a helium
abundance at least 2--3 times solar, or it is a newly-discovered DAB or
DBA WD. Observations in the blue part of the optical spectrum will be
required to resolve the ambiguity.

\noindent {\em 1101+364, 1123+189}. The PG coordinates for these
objects (RA 11:01:35.3, DEC +36:26:20; RA 11:23:46.6, DEC +18:55:18)
are in error. The objects are located at RA 11:01:46.3, DEC +36:26:56
and RA 11:23:42, DEC +18:55:50, respectively (1950).

\noindent {\em 1104+602}. This object (EG 75) is mistakenly identified
in the PG survey as PG 1100+604. The correct coordinates are RA
11:04:43, DEC +60:14:48 (1950). The spectrum of the object found at
the PG coodinates (RA 11:00:43.5, DEC +60:27:41) and agreeing in the
telescope guide star camera with the position on the PG finder plate
resembles that of a main sequence G star, based on the depth and
narrowness of the \halpha\ line.

\noindent {\em 1204+451, 1459+305}. These objects are visual
binaries. The DA WDs are the western (1204+451) and southern
(1459+305) stars.

\noindent{\em 1422+095}. This object is the well-studied pulsating DA
GD 165A (\cite{gd165}), which has a very low-mass (possibly
sub-stellar), distant companion (\cite{zuckbeck}). We detect the WD as
a velocity variable (see Table \ref{ddvariables}). If the cooling age
of the older WD secondary could be determined, an improved lower limit
could be placed on the cooling age of the potential brown dwarf
companion. Analysis of delays in the WD pulsation arrival times would
be another way to study the system's orbital properties.

\noindent {\em 1411+219}. This object, identified as composite in the
PG survey, is a visual double, which in retrospect can be seen to be
barely resolved upon close inspection of the PG finder plate. The DB
WD is the northern component of the pair and shows weak \halpha\
absorption in addition to strong He {\sc i} \lam{6678}. The
southern component shows late K- or M-type spectral features.

\noindent {\em 1559+369, 1600+369}. These objects are separate entries
in the catalog of \cite{mccook}, but in fact are the same object. The
PG coordinates (RA 16:00:05.2, DEC +36:55:29) are in error. The
object is located at RA 15:59:33, DEC +36:56:58 (1950).

\noindent {\em 1713+332}. This high-proper motion star has 1996.3
coordinates RA 17:13:46.7, DEC +33:16:22. 

\noindent {\em 2009+622}. This object was known to show an infrared
excess and was spectroscopically observed by Schultz, Zuckerman, \&
Becklin (1996) in an effort to detect \halpha\ emission from the
inferred low-mass main sequence companion, with ambiguous results. We
clearly detect strong, variable \halpha\ emission in the core of the
DA WD absorption line. In addition, the emission velocity is seen to
vary by some 270 \kms, indicative of a very short-period system ($P$
\lta 1 day).

\noindent {\em 2126+734}. This object was found to have a cool WD
companion at a separation of 1\farcs 4 by Zuckerman et al. (1997).

\section{Summary of Known DD Systems}

In Table \ref{knowndd}, we present all known DDs in close
(short-period) binary systems having measured orbital periods and
primary masses, including the 5 recently discovered sdB+WD pairs with
confirmed periods. We omit the ultra-short period AM CVn systems,
since they are likely not detached DDs, and their primary stars are
believed to be currently accreting from He secondaries. In any case,
their primary masses cannot be directly measured, and there is a
considerable range in the masses implied by the theory of their
evolution (e.g., \cite{bwarner}).  Successive table columns give the
object, its name, the period in days, logarithm of the period, an
estimate of the mass of the primary component, total system mass when
known, and the source of the data. We now wish to compare the results
of the observations with the predictions of theory.

\section{Predictions of Binary Evolution Theory}

Population synthesis calculations of the binary population of the
Galaxy (YLTS; ITY) predict the birthrate and observable number
distributions of DD and WD+MS pairs. Here, we present the results of a
recent set of calculations, which like ITY and YLTS take into account
the effects of observational selection due to white dwarf cooling and
the relative brightness of the components, and which also allow for a
more complete comparison with observations (see below). We note that
in the context of DD evolution, the primary component is taken to be
the visible member of the binary, while in fact the secondary (dimmer)
component was most often the originally more massive star. Exceptions
are Algol-like systems where conservative mass transfer has reversed
the original mass ratio.

The observable (younger and brighter) component may be a helium (He)
or carbon-oxygen (CO) WD, while the older, dimmer component (usually
unobserved) may be a He or CO WD, or a low-mass main sequence star. In
systems composed of two close WDs, the theory predicts that 82\% of
the primaries are He WDs, in good agreement with observations. In
fact, Table \ref{knowndd} shows that of all known DDs with confirmed
periods and primary masses, every WD except L~870--2 is inferred to be
of the He variety based on its low mass ($<$ 0.5 \msun). The five sdB
primaries are expected to be He stars which eventually will evolve to
lower-mass CO WDs after cessation of core helium-burning.

The goal of this and previous surveys aimed at detecting DD systems,
beyond the desire to identify a pre-\snia\ system, is to confront the
observations with the mass and period distributions predicted by the
theory. Previously, these distributions have been considered
independently (cf. Figures 1 and 2 of YLTS; Figure 2 of ITY). Here, we
consider the bivariate distribution $N$(log $P$, log $M$), which may
provide insight into different evolutionary channels that, while
producing DD systems with similar periods, might produce systems with
significantly different masses.

Using the same population synthesis code as ITY, we have recalculated
the numbers of DD and WD+MS pairs on a fine log $P$ vs. log $M$
grid. The grid range and spacings are \logm\ from --0.8 to 0.15 in
steps of 0.025, and log $P$ (in days) from --1.5 to 4.5 in steps of
0.1. The masses are those of the primary (visible) WD. In Figure
\ref{bivary}, we display a halftone plot of the theoretical DD number
distribution, with the greyscale spanning values between 0 and
\scinote{1.25}{5}.  The filled symbols represent all known DDs in
Table 3, with circles for WD primaries and diamonds for sdB
primaries. Masses for the WD primaries have been taken from the
literature. For the sdB stars, a mass of 0.5 \msun\ has been assumed,
based on their location in the temperature--gravity plane (\cite
{sbkl}; \cite{rasjwl}).

Remarkable structure is evident in the bivariate distribution which is
obscured in the projected univariate distributions $N$(log $P$) and
$N$(log $M$). A strong ridgeline extends from the low-mass,
short-period region toward higher masses and longer periods. The
structure is understood as the DD ``birthline'' -- at a given mass
there is a period threshold beyond which systems do not interact and
never evolve through a common envelope phase. The numbers of systems
are largest at low mass, where the initial mass function produces the
largest numbers of progenitor stars. The ``island'' at \logm\ = --0.2
represents the WDs with CO cores. We further discuss the comparison
between the theory and observations in Section 8.

\section{Selection Effects and Detection Efficiency}

Our survey method fails to detect all bonafide short-period binaries
for several reasons:
\begin{enumerate}
\item 
Random inclination of the orbital planes to the line of sight masks
radial velocity variations for near plane-on systems.
\item 
Our temporal sampling imposes two different constraints: $i$) Exposure
times for most objects were between 15 and 30 minutes, which would
cause smearing of the observed line profiles for orbital periods
shorter than 1--2 hours. This affects all objects in the same way. We
adopt a period of 2 hours as a conservative lower limit to our
sensitivity bandpass.  $ii$) Some orbital periods will have low
sensitivity due to period aliasing, affecting objects differently
depending on the exact sequence of observation times.
\item
In systems with low-mass secondaries, the orbital velocity of the
primary will be lower than for high-mass secondaries. WD+MS pairs are
especially afflicted with this bias since the MS companion typically
must have a mass $\leq$ 0.3 \msun\ in order to have remained hidden
until detected spectroscopically. However, this bias is mitigated by
the frequent presence of reprocessed \halpha\ emission in the
MS secondary's photosphere, which is easily detected.
\item
Given our modest radial velocity precision (20--30 \kms\ at our
intermediate spectral resolution and S/N ratio), very long period
systems will have velocities too small to detect variations.
\item
In very hot and/or strongly magnetic DA and DAO WDs, the line
profiles are usually shallow and very broad, and they lack the sharp,
non-LTE absorption core of line profiles of cooler or non-magnetic
WDs. Velocity differences are therefore very difficult to discern,
except for the very hottest DA or DAO WDs, like Feige 55, which often
show narrow photospheric emission in the \halpha\ absorption line
(cf. Holberg et al. 1995). Fortunately, few WDs in the sample fall into
this category.
\end{enumerate}
Previous investigators (RS, FWG, \cite{tmarsh1}, \cite{tmarsh2}) have
used a number of analytical methods for estimating detection
probabilities. Each approach attempted to address a particular
statistical question. For example, FWG calculated the probability that
no binaries at all were found in their sample, while Marsh et
al. (1995) calculated the ratio of the probabilities that individual
objects were binaries vs. single stars. Our own concern is to
estimate the fraction of bonafide DD systems our survey fails to
detect due to selection effects. We choose a Monte Carlo approach,
beginning with our calculated mass/period distribution (Figure
\ref{bivary}).

For each object in Table \ref{wdsample}, we first randomly select an
orbital period and primary mass from the theoretical distribution
using the rejection method (\cite{pressetal}). We deem the experiment
to have failed and continue with the next object if the orbital period
is less than 2 hours.  We then randomly select a secondary mass, using
the same procedure, from one of the two distributions for WD+WD given
in the top panel of Figure 4 of ITY, according to whether the primary
is a He or CO WD. The orbital separation and primary WD velocity are
computed, an orbital plane inclination is chosen from the
probability distribution $P(i) = \sin i$, and an initial orbital phase
is chosen from a uniform deviate. Orbital phase zero is defined as
the positive-going crossing of the systemic velocity by the primary's
velocity curve.

With all relevant orbital parameters fixed, radial velocities are then
computed for each mid-exposure time of observation, and the maximum
radial velocity difference is determined. We consider the simulated
system to have been detected if the maximum velocity difference
exceeds 65 \kms, or 2.5 times an assumed 1$\sigma$ precision of 25
\kms\ (consistent with our definition of definite detection when the
observed line core variation exceeds 1.5\AA). A ``hit'' is recorded
for each detected object, and the next object is then tested.

The experiment was repeated for all WDs in Table \ref{wdsample}, and
for each complete pass through the sample the detection probability
was taken to be the fraction of hits. This is specific to our sample,
as it depends on the sequence of observation times for each object,
and it assumes a parent population of 100\% WD+WD binaries with the
theoretical log $P$--log $M$ distribution of Figure \ref{bivary}. The
procedure was repeated 1000 times each for two cases of orbital plane
inclination: $i = 90^\circ$ (edge-on) and $P(i) = \sin i$, and
for three cases of allowed periods: all periods in our calculated
grid, $P$ between 2$^{\rm h}$--20$^{\rm d}$, and $P$ between 2$^{\rm
h}$--10$^{\rm d}$.

The result for the most favorable condition, with $i = 90^\circ$ and
$P$ between 2$^{\rm h}$--10$^{\rm d}$, is shown in Figure
\ref{ddprob}, where the distribution of detection efficiencies is
nearly gaussian with a mean of 83\% and standard deviation of 3\%. The
remaining results are given in Table \ref{allddprob}, which shows that
the effects of random inclination of orbital planes and low velocities
associated with longer orbital periods are both important contributors
to decreased detection efficiency. We adopt an upper limit of 10--20
days to our sensitivity bandpass (with a lower limit of 2 hours),
corresponding to a detection efficiency of \twid 70\% over our entire
sample. We do not consider the probability of duplicity for individual
objects, as theoretical predictions of the fraction of DDs and the
mass and period distributions pertain to the entire population of WDs
and sdBs in the Galaxy. Our method of estimating detection
efficiencies allows an evaluation of selection effects on the raw
observed fraction of DDs detected in the sample as a whole.

\section{Comparisons with Theoretical Predictions}

Comparison of the existing observations with the bivariate
distribution of Figure \ref{bivary} is not straightforward for two
reasons: 1) The number of DD systems discovered to date which have
orbital periods and primary (visible) masses determined is still quite
small. 2) The measurement errors for the WD and sdB masses, based 
primarily on surface gravity errors, are generally \twid 0.05--0.10 in
\logm\ and do not permit an unambiguous comparision. Of the six
systems near \logm\ = --0.3 and lying just below the CO peak at \logm\
\twid --0.2 (\twid 0.6 \msun) in Figure \ref{bivary}, five have sdB
primaries with masses near 0.5 \msun. As such, they sensibly ought to
lie slightly below the CO peak, and we do not consider this a serious
disagreement. The sdB periods do agree very well with the predicted CO
WD period distribution.

Four objects lie comfortably near either the DD birthline or the two
peaks at \logm\ = --0.5 and --0.65. The remaining objects do not
appear to lie near peaks in the distribution. The agreement with
theory would be improved if the derived WD primary masses were
systematically too large, or if the theoretically predicted masses
were systematically too small. We note that the effect of varying the
common envelope efficiency parameter \ace\ (which was taken to be
unity in our calculations) is largely to shift the predicted periods,
not the primary masses (cf. \cite{ibenlivio}). 

It is possible that at least some observed DD primary masses are
overestimated due to the presence of a fainter and usually more
massive WD companion. Line profile analysis of the composite optical
spectrum of the brighter, lower-gravity visible WD and a dimmer, but
higher-gravity WD might result in a derived surface gravity higher
than that of the primary alone and would imply a too-large primary
mass. On the other hand, in cases where the mass ratios have been
measured (L~870--2, 1101+364, 0957--666), the WDs have similar masses,
and it is unlikely that companions contribute more than 20\% of the
light in 1241--010, 1317+453 and 1713+332 (T. Marsh, private
communication). Also, the radial velocity data for 0957--666 almost
certainly preclude reducing the WD masses.

There may be another effect which distorts the bivariate number
distribution in the $\log P$--$\log M$\ diagram. The theoretical grid
was constructed assuming that all WDs follow the same cooling sequence
computed by \cite{ibentut84b} for 0.6 $M_\odot$\ WDs. Following their
calculations, we assumed that all WDs are detectable for a time $t =
\min (10^8\,{\rm yr, time~to~merger}).$~ In other words, we assumed
that the luminosity of the white dwarf decreases below a certain
detection limit in $10^8$\,yr, irrespective of its mass. Until very
recently, cooling sequences for low mass WDs were not available.
However, recent calculations (\cite{hansphinn}) for $M$ = 0.15--0.45
\msun\ show that the more massive WDs (\twid 0.45 \msun) cool to the
same luminosity several times slower than the lower-mass ones (\twid
0.15 \msun). This may be understood as an effect of the larger surface
areas of low-mass WDs. An attempt to quantify the effect of this mass
dependence of cooling curves on the number distributions of WDs is
currently underway (Tutukov \& Yungelson, in preparation). The
difference in cooling curves may shift the theoretical
distribution in Figure 3 more in favor of high-mass systems, and the
strong concentration of WDs along the "birthline" below \logm\ = --0.6
may diminish.

In light of the difficulty still associated with interpreting the
theoretical bivariate distribution, the total predicted number of DD
systems and the projected log $P$ and log $M$ distributions remain at
present the best means of comparing the theory and observations. In
Table \ref{ddtheory}, we present the predicted fractions of DD and
WD+MS systems for three values of \ace. Successive columns give \ace,
the number of close DDs vs. all WDs (i.e., the sum of all single WDs
plus all wide and close pairs), the number of pre-\snia\ systems
vs. all WDs, and the number of close WD+MS pairs vs. all WDs. The
pre-\snia\ column describes DDs which have a total system mass greater
than \mch\ and may merge within a Hubble time. The table also
illustrates how many WDs have to be surveyed for a detection of a
pre-\snia, and how much more effort has to be invested before
observations will better constrain the theory.

We now compare the predicted numbers of DDs with two observed samples,
the one reported here and that of BGRD, as these have similar
magnitude limits and are sensitive to the same range of orbital
periods (\twid 2 hr to a few days). Three comparisons are possible:
\begin{enumerate}
\item 
BGRD found 1 DD and 4 candidate DDs in their sample of 54 WDs (\twid
1/11), and 4 WD+MS pairs, 2 which were unresolved, and 2 which were
resolved as visual binaries (\twid 1/13). In our combined sample of
107 WDs and 46 sdBs we find 6 confirmed DDs and 15 candidate DDs, not
all of which are new detections. We note that seven of these have sdB
primaries, which have core helium-burning lifetimes of \scinote{1--2}{8}
years, comparable to the interval during which WDs are the brightest
and are most likely to be discovered (cf. ITY). However, the field sdB
stars have a birthrate \twid 0.5\% of that of field WDs
(\cite{rasjwl}), and they are thus quite heavily over-represented in
our sample. Considered separately, the fractions of DD systems with WD
and sdB primaries among all stars in our sample are \twid 1/8 and
1/7, respectively. Including the visual doubles and PG composites
noted in Tables \ref{wdsample} and \ref{sdbsample}, 12 WD+MS pairs are
found in our sample (\twid 1/13). Excluding duplication of objects in
common, the fractions of DDs and WD+MS pairs found among all WDs in
our and BGRD's combined samples are 1/9 and 1/12, respectively.
These results are in good agreement with columns 2 and 4 of Table
\ref{ddtheory} for values of \ace\ near 1.0 and tend to exclude much
lower values of \ace.
\item
The population synthesis code also predicts the expected projected
period and mass distributions of close DDs. We first compare the
observations with the predicted period distribution in Figure
\ref{ddnlogp}. The theoretical prediction is the solid histogram
(taken as a projection onto the log $P$ axis from our most recent
calculation of the bivariate distribution), and the dot-dashed
histogram is the observed period distribution of the DDs in Table
\ref{knowndd}, including the newly-discovered sdB+WD pairs. The
observed distribution (arbitrarily normalized) is generally in
agreement with the predicted distribution, albeit possibly offset to
shorter periods. Part of the apparent deficit of systems at longer
periods is due to the decline in our detection efficiency at orbital
periods longer than 10--20 days. 
\item 
In Figure \ref{ddnlogm}, the predicted distribution of primary masses
of close DDs is given as the solid histogram. There is a broad peak
near \logm\ \twid --0.5, corresponding to He WD primaries, and a
second strong, narrow peak at \logm\ \twid --0.2, corresponding to CO
WD primaries, with a tail extending to masses in excess of 1.2
\msun. The dot-dashed histogram is the observed mass distribution of
the DDs in Table \ref{knowndd}. The observed distribution is generally
in agreement with the predictions of theory, except for the apparent
excess of stars in the bin at \logm\ = --0.3. All but one of these
(L~870--2) are sdB stars, which as we have previously noted, are
over-represented in our sample.

The theory also predicts that some 18\% of primary stars in DDs have
CO cores. We therefore expect to find 2 CO WDs in the confirmed
short-period DD sample; instead, we find none. We note that of the
radial velocity variables in Table \ref{ddvariables}, 10 are in common
with the sample of \cite{bsl}, among which only three objects have
derived masses $\geq$ 0.5 \msun: 1232+479 (0.53 \msun, weight 1),
1446+286 (0.82 \msun, weight 2), and 2117+539 (0.50 \msun, weight
1). Four other objects are in common with the sample of Bragaglia et
al. (1995), but only one has a derived mass $\geq$ 0.5 \msun: 1422+095
(0.51 \msun; weight 1). It is crucial to obtain accurate masses for
the remaining objects in Table \ref{ddvariables} to identify any CO
primaries which may reside in potential pre-\snia\ systems. The
possibility that the primary stars' optical spectra may be
diluted with that of dimmer, possibly higher-gravity companions must
be taken into account.
\end{enumerate}
When confirmed, the 18 new candidate DDs (including the sdB variables)
from our survey will increase the number of known short-period DDs to
29 systems, approaching the number predicted to include a super-\mch\
system with a period short enough to merge in a Hubble time (\twid
1/40 for \ace\ = 1.0; see Table \ref{ddtheory}). Clearly, in addition to
the follow-up observations to determine orbital periods and masses of
the DD candidates in the current sample, further survey work is
required in order to expand the WD sample by approximately a factor of
2 (for \ace\ of 1.0).

The previous discussions have not taken into account DD systems missed
in our survey due to selection effects. We first consider the systems
with WD primaries. Assuming an efficiency of 70\%, our raw number of
14 confirmed and candidate DD systems would imply the existence of
some 20 bonafide DD systems with WD primaries in our sample of 107
WDs, or a fraction just under 1/5 of all WDs. The same result is
obtained from the sample of DDs with sdB primaries. Table
\ref{ddtheory} shows that this substantially exceeds the predicted
fraction for \ace\ \twid 1. Such a high implied fraction of close
binaries among all WDs suggests that, formally, \ace\ $>$ 1--2. We
note that the condition that \ace\ $>$ 1 has been suggested as a
requirment for an explanation of the correlation of orbital periods
and eccentricities of high mass binary pulsars (e.g., \cite{zwart})
and millisecond binary pulsars (e.g., \cite{tauris1}; \cite{tauris2};
\cite{vdheuvel}). However, since detailed numerical models of the CE
process, which involve energy sources other than gravitational (e.g.,
\cite{livio89}) are not available, any treatment of the evolution of
orbital separations inside the CE, which ultimately determines the
predicted number distributions, is at best an approximate estimate.

\section{Binarity Among DAO White Dwarfs}

The DAO WDs are a small class of hot, hydrogen-rich stars that show
weak, narrow \ion{He}{2} \lam{4686} absorption in addition to the
broad Balmer lines. \cite{dao94} analyzed 12 DAOs and found that half
of them have such low surface gravities (and by implication low
masses), that unlike most WDs they cannot be products of post-AGB
evolution. \cite{dao94} suggested that the low-mass DAOs might be the
product of post-extended horizontal branch (post-EHB)
evolution. Another alternative was raised by \cite{feige55}, who noted
that the very low mass of the DAO Feige 55 ($<$ 0.4 \msun) might be
related to its double degenerate nature. The EHB stars (sdBs and the
hottest BHB stars) are thought to have core masses very near 0.5
\msun, and post-EHB evolution alone is unlikely to have produced Feige
55's low core mass. In contrast, close binary evolution is
predicted to produce a wealth of low-mass remnants, as observations
confirm.

Feige 55 was detected as a radial velocity variable in IUE observations,
and its orbital parameters were subsequently determined from data
including observations obtained during our 1995 March observing run
(Holberg et al. 1995). We also observed two other DAO WDs, 1210+533 and
2342+806. The former star has relatively weak \halpha\ absorption and
has neither a sharp non-LTE absorption core nor a central reversal. It
does not appear to be a velocity variable, but we cannot rule it out
conclusively. The latter star (GD 561) is a near-twin of Feige 55, with
a somewhat hotter temperature and an even lower surface gravity and 
mass. Our observations do show a distinct emission reversal
like that of Feige 55, but the star is constant in velocity at a level
\lta 15 \kms\ over a span of 6 nights. These results point out the need
for further investigation into the possible connection between close
binary evolution and the very low masses of approximately half of all
DOA WDs.

\section{Conclusions}

We have performed a radial velocity survey of a total of 153 field WDs
and sdBs, most of which were not previously known to be binary. In the
combined sample, we have discovered 18 new DD candidates with WD or
sdB primaries, and 1 new confirmed WD+MS pair. Among the 7 sdB
variables, we have obtained orbital solutions for the short-period
sdB+WD pairs Feige 36 ($P$ = 8\fh 5) and Ton 245 ($P$ = 2\fd 5), and
Moran et al. (1998, in preparation) have found orbital periods for
0101+039 ($P$ = 16\fh 1 or 13\fh 7), 1432+159 ($P$ = 5\fh 39), and
2345+318 ($P$ = 5\fh 78). Our conclusions are:
\begin{enumerate}
\item 
The population of close, short-period DD systems predicted by the
theory of close binary star evolution does in fact exist in
significant numbers.  The raw observed fractions of confirmed
short-period DDs and WD+MS pairs among all WDs are in satisfactory
agreement with the predictions of theory for values of the common
envelope efficiency parameter \ace\ near 1.0. However, applying a
correction for selection effects implies an excess of existing
DDs over the predicted number. A resolution of this apparent
discrepancy will have to await a more complete understanding of the CE
process.
\item
The observed DD orbital period and primary mass distributions do fall
very nearly in the peak of the predicted distributions for \ace\ =
1.0, although possibly shifted slightly. We note that the discrepancy
of the predicted and observed period distributions could be made to
vanish by a small decrease in the value of \ace, a free parameter in
the theory (``the parameter of ignorance'', I. Iben). The discrepancy
with the predicted mass distribution is less amenable to fine-tuning
of the theory, and we suggest that some previous primary WD mass
determinations may have been systematically biased to higher masses
due to dilution of the optical spectra by the presence of unseen,
higher-gravity companions. Over-representation of sdB primaries in the
combined sample is responsible for the apparent excess of stars with
masses near 0.5 \msun. A refined analysis of the effect of the
mass-dependence of WD cooling sequences upon the $\log P$--$\log M$
number distribution is also required.
\item 
The low-mass systems observed by Marsh et al. (1995) and Marsh (1995)
tend to have short orbital periods, in agreement with theoretical
predictions. However, their total system masses likely are too small
to make them viable \snia\ progenitors. Furthermore, the primary stars
must have He cores, while CO white dwarfs are thought to be the best
candidates to undergo the central ignition and detonation of carbon
required to produce a \snia. \\
\item The one known super-\mch\ DD system, LB~11146, has an unknown
orbital period, although its projected orbital separation is \lta 1
a.u. The large remnant masses suggest that the progenitor stars also
were massive, had large radii in the AGB phase, and that the system
might have gone through at least one CE phase leading to a short
orbital period. However, no ``loaded gun'' has yet been found, i.e., a
super-\mch\ system with $P$ \lta 10$^{\rm h}$. On the other hand, the
sample of confirmed short-period DDs is still smaller than the number
predicted to contain a massive, short-period system. When confirmed,
the 18 new DD candidates discovered in our survey will increase the
number of known short-period DDs among all WDs to a fraction
approaching that which theory predicts will include a pre-\snia\
system. \\
\end{enumerate}
We shall continue with follow-up observations to solve for the orbital
parameters of the new WD+WD and sdB+WD candidates found in our survey,
and with the survey work itself to further expand the
sample. Additional work on the DAO WDs is needed to elucidate the
connection between close binary evolution and the very low masses of
half of DAOs. Infrared observations, especially for the
higher-luminosity sdB stars, will be required to distinguish main
sequence companions having masses as low as \twid 0.1 \msun\ from
compact companions.  

Our survey has revealed a significant new population of close DD
systems. Even if no convincing potential \snia\ progenitor is found,
the distributions of DD orbital periods and primary masses (and
secondary masses and mass ratios, where observeable) constitute
powerful probes of the CE process, and a meaningful test of the
predictions of the theory of close binary evolution now seems within
reach.

\bigskip

Acknowledgements: This work has been supported by NASA Grants
NAGW-2678, G005.44000, and by the Director's Discretionary Research
Fund at the Space Telescope Science Institute. LRY acknowledges
support through RFBR grant 960216351, and is grateful for the
hospitality of Meudon Observatory. RAS acknowledges fruitful
conversations with Tom Marsh, Jim Liebert, Jay Holberg, and Gary
Schmidt.  LRY acknowledges day-to-day discussions of stellar evolution
with Alexander Tutukov. The authors are grateful for the long-term
committment of the KPNO TAC and awards of valuable Kitt Peak 2.1-m
time, crucial to the success of our large-scale survey. We thank the
anonymous referee for a critical review of the manuscript.

\clearpage

\figcaption[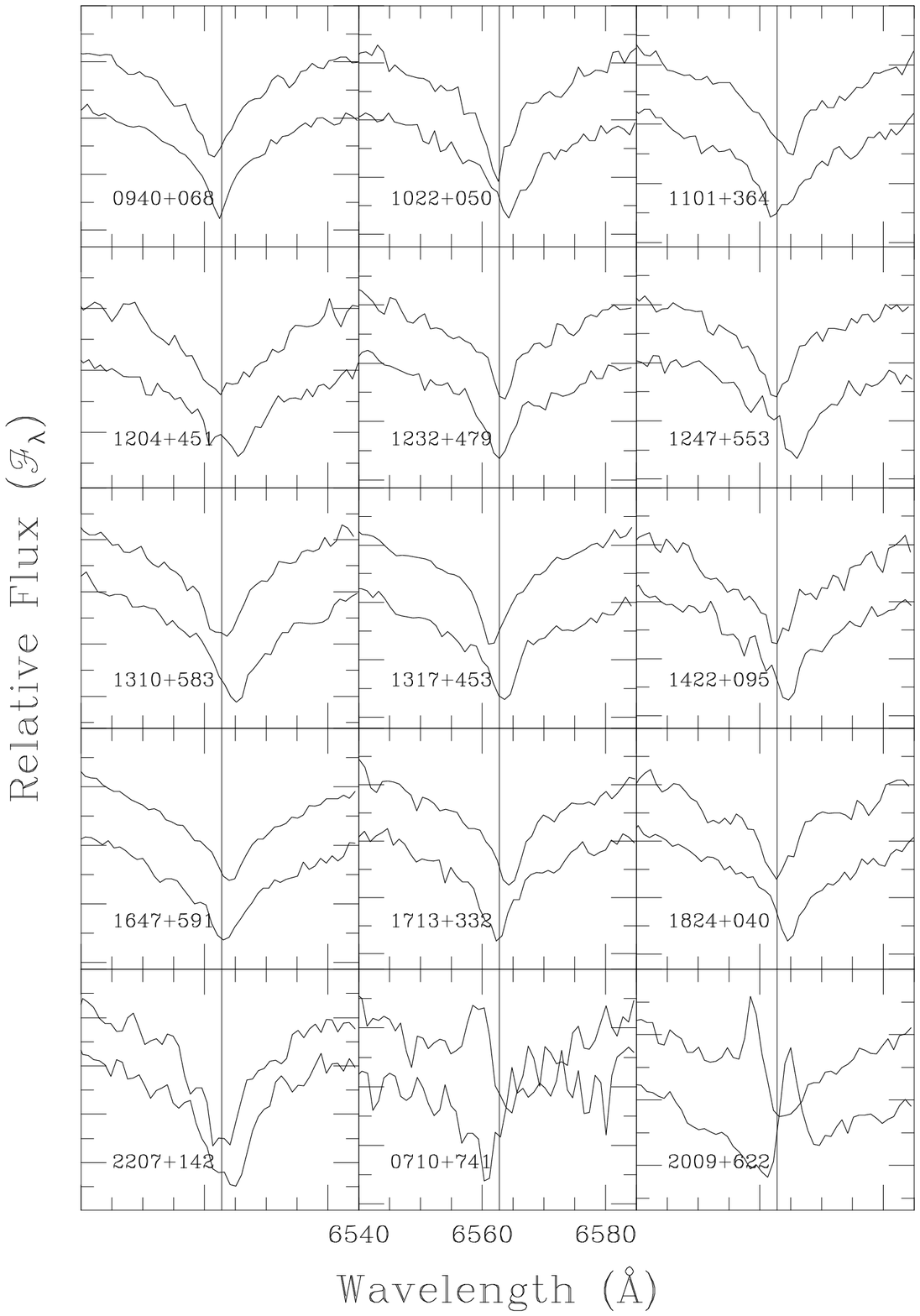]{Radial velocity variability at \halpha. The
first 13 panels show spectra of WD radial velocity variables. The
laboratory wavelength of \halpha\ is marked with the vertical
line. The sdB variables of Saffer (1991) are excluded. The
two panels at bottom right show narrow, strongly variable \halpha\
emission in the cores of WD+MS pairs. \label{multistack}}

\figcaption[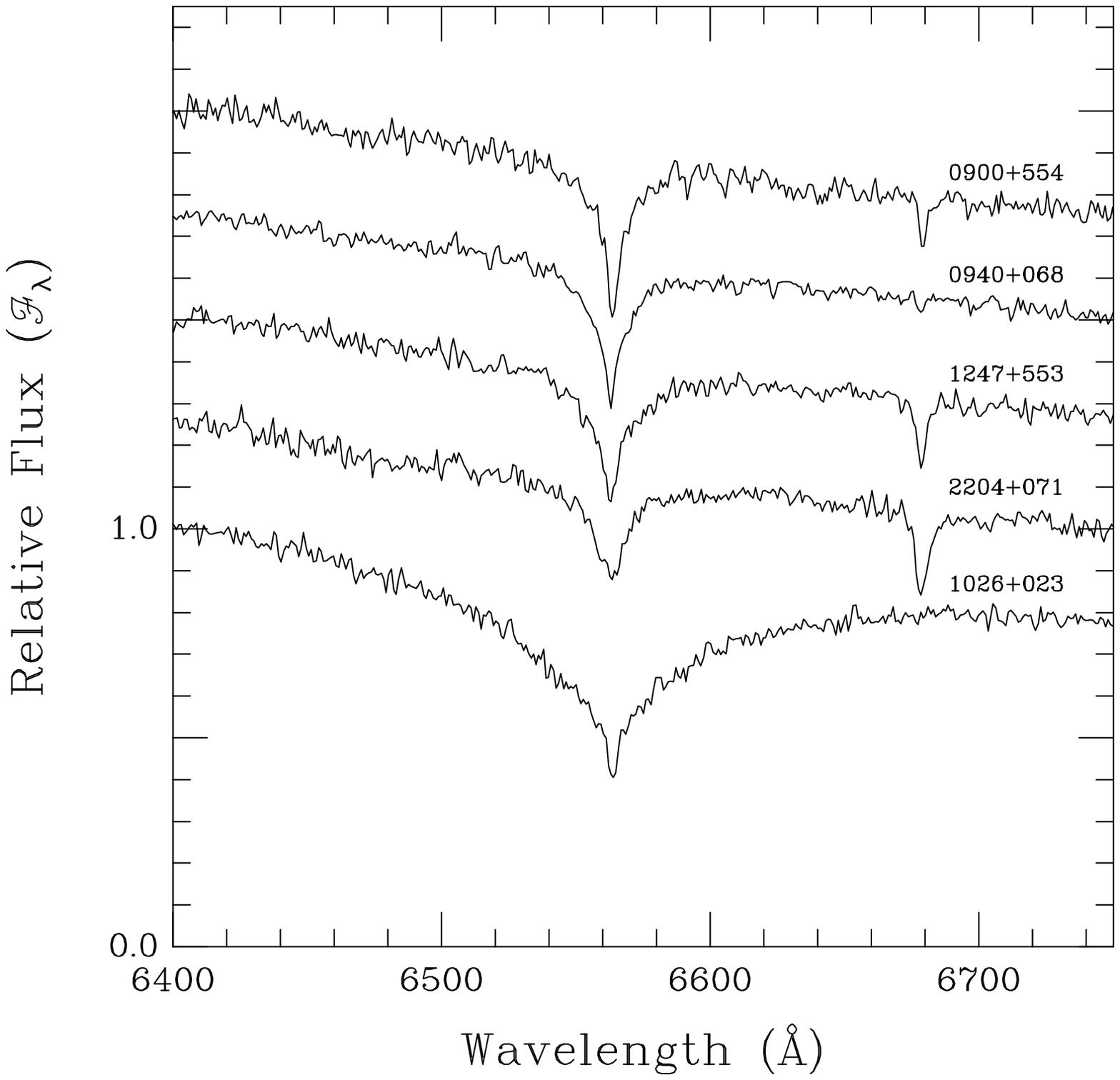]{Narrow, sharp \halpha\ and He {\sc i}
\lam{6678} absorption identify the first 3 spectra as belonging to
likely BHB stars (0900+554 and 1247+553) or sdB stars (0940+068). The
fourth spectrum shows a probable new DAB or DBA star (see
the text for a resolution of ambiguity of the PG identifier). The
bottom spectrum shows the highly Stark-broadened profile of a normal
DA WD for comparison. \label{wd2sdb}}

\figcaption[nlogpnlogm.ps]{The predicted observable number
distribution of DD systems as a function of mass and orbital
period. The masses are those of the primary (visible) star. The
greyscale spans values between 0 and \scinote{1.25}{5}. The strong
ridgeline extending up and to the right is the DD ``birthline'', below
which binary systems never interact and experience a common envelope
phase. DD systems with known orbital periods and primary masses are
plotted with their mass errors. The filled circles and diamonds denote
WD and sdB primaries, respectively. \label{bivary}}

\figcaption[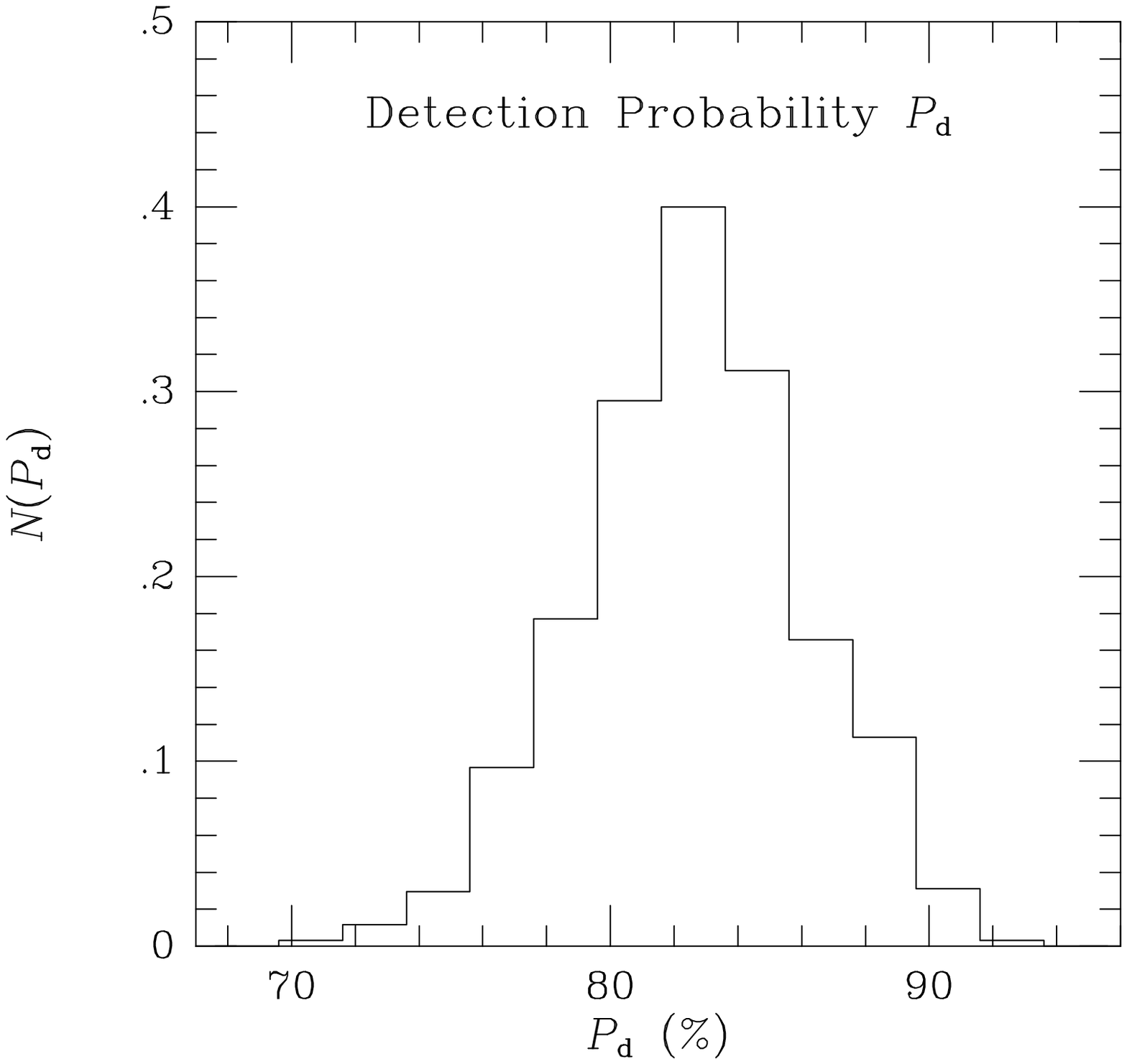]{Distribution of detection probabilites for the
WD sample computed from 1000 trials, assuming the most favorable
condition. The orbital plane is fixed at an inclination $i = 90^\circ$,
and orbital periods are restricted to the range 2$^{\rm h}$--10$^{\rm
d}$. The distribution is centered on a detection efficiency of 83\%
with a standard deviation of 3\%. The integral under the curve is
unity. See the text for details. \label{ddprob}}

\figcaption[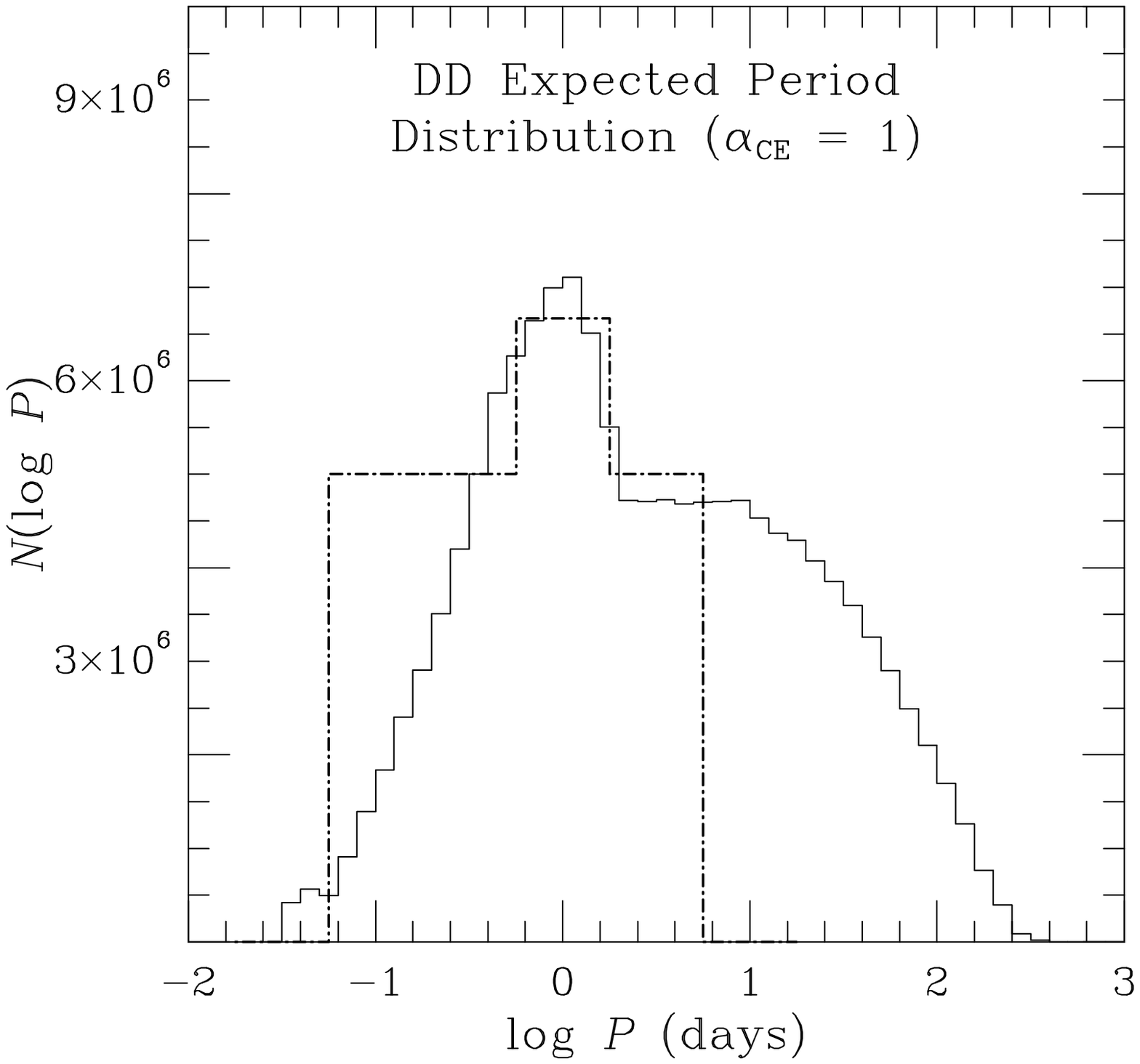]{Comparison of the predicted and observed
projected DD period distributions. The heavy dot-dashed histogram is
the observed distribution, arbitrarily normalized. \label{ddnlogp}}

\figcaption[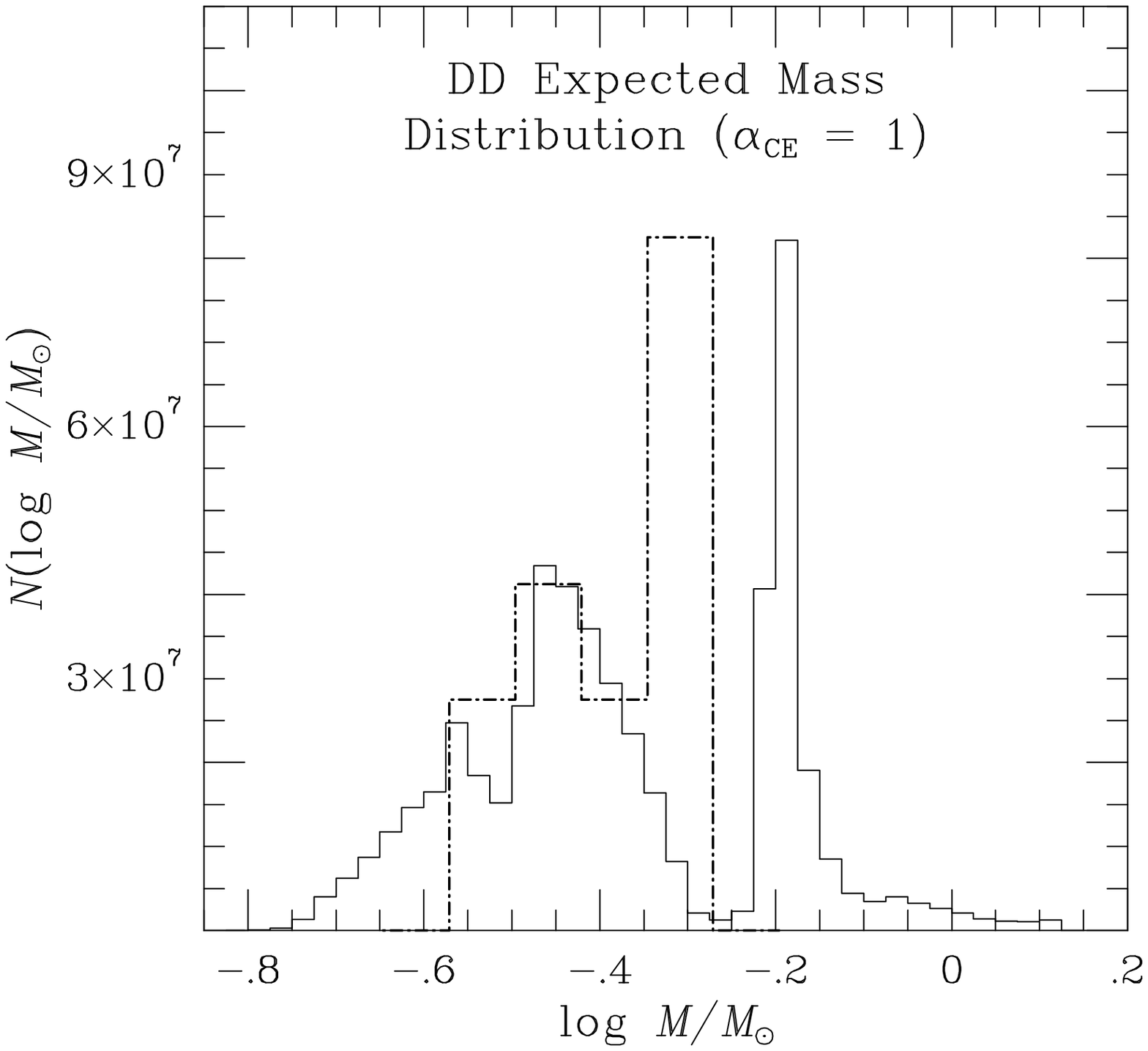]{Comparison of the predicted and observed
projected DD mass distributions. The heavy dot-dashed histogram is the
observed distribution, arbitrarily normalized. \label{ddnlogm}}

\singlespacing

\begin{deluxetable}{cllccl}
\footnotesize
\tablewidth{400pt}
\tablecaption{White Dwarf Sample \label{wdsample}}
\tablehead{
\colhead{WD}         & \colhead{Name}     & \colhead{$V$}   & 
\colhead{$B$--$V$}   & \colhead{$U$--$B$} & \colhead{Reference} 
}
\startdata
0002$+$729 & GD 408       & 14.35 & $-$0.08 & $-$0.91 & 1,13     \nl 
0004$+$330 & GD 2         & 13.82 & $-$0.29 & $-$1.21 & 1,2      \nl 
0101$+$039 & Feige 11     & 12.06 & $-$0.24 & $-$0.98 & 3        \nl 
0101$+$048 & G 1$-$45     & 13.96 & $+$0.31 & $-$0.57 &          \nl 
\medskip                                                             
0112$+$142 & PG           & 15.0: & $-$0.59 & $-$0.47 & 3,4      \nl 

0134$+$833 & GD 419       & 13.08 & $-$0.64 & $-$0.06 & 4        \nl 
0148$+$467 & GD 279       & 12.44 & $+$0.06 & $-$0.63 & 1,2      \nl 
0151$+$017 & G 73$-$4     & 15.00 & $+$0.16 & $-$0.59 &          \nl 
0205$+$250 & G 35$-$29    & 13.22 & $-$0.04 & $-$0.85 & 2        \nl 
\medskip                                                             
0208$+$396 & G 74$-$7     & 14.50 & $+$0.30 & $-$0.47 &          \nl 

0232$+$525 & G 174$-$5    & 13.75 & $-$0.55 & $+$0.08 & 4        \nl 
0401$+$250 & G 8$-$8      & 13.81 & $+$0.11 & $-$0.58 & 1        \nl 
0425$+$168 & LP 415$-$415 & 14.02 & $-$0.09 & $-$0.97 &          \nl 
0437$+$138 & LP 475$-$242 & 14.86 & $-$0.43 & $-$0.25 & 4,13     \nl 
\medskip                                                             
0439$+$466 & LSV $+$46 21 & 12.67 & $-$0.30 & $-$1.24 & 5        \nl 

0453$+$418 & GD 64        & 13.77 & $+$0.23 & $-$0.61 & 1        \nl 
0532$+$414 & GD 69        & 14.75 & $+$0.32 & $-$0.54 &          \nl 
0549$+$158 & GD 71        & 13.06 & $-$0.25 & $-$1.16 & 1        \nl 
0553$+$053 & G 99$-$47    & 14.08 & $+$0.61 & $-$0.11 & 6        \nl 
\medskip                                                             
0612$+$177 & G 104$-$27   & 13.40 & $-$0.15 & $-$0.98 & 1,2,7,8  \nl 

0637$+$478 & GD 77        & 14.80 & $+$0.13 & $-$0.64 & 6        \nl 
0644$+$376 & G 87$-$7     & 12.10 & $-$0.08 & $-$0.92 & 1,2      \nl 
0651$-$020 & GD 80        & 14.82 & $-$0.21 & $-$1.19 & 7        \nl 
0706$+$295 & KUV          & 15.45 & $+$0.09 & $-$0.66 &          \nl 
\medskip                                                             
0710$+$741 & GD 448       & 14.97 & $-$0.06 & $-$0.75 & 9        \nl 

0713$+$584 & GD 294       & 12.04 & $+$0.00 & $+$1.57 & 1,10     \nl 
0743$+$442 & GD 89        & 14.92 & $+$0.07 & $-$0.73 &          \nl 
0834$+$501 & PG           & 15.82 & $-$0.36 & $-$1.20 & 5        \nl 
0839$+$232 & PG           & 14.42 & $-$0.14 & $-$0.98 & 1        \nl 
\medskip                                                             
0854$+$404 & GD 98        & 14.87 & $-$0.13 & $-$0.94 &          \nl

0900$+$554 & PG           & 13.84 & $-$0.05 & $+$0.49 & 1,10,11  \nl 
0938$+$550 & PG           & 14.77 & $-$0.52 & $+$0.05 & 4        \nl 
0939$+$262 & Ton 21       & 14.53 & $-$0.35 & $-$1.16 &          \nl 
0940$+$068 & PG           & 13.70 & $-$0.19 & $-$1.03 & 7,11,20  \nl 
\medskip                                                             
0943$+$441 & G 116$-$52   & 13.32 & $+$0.07 & $-$0.54 & 1        \nl 

0945$+$246 & LB 11146     & 14.32 & $-$0.47 & $-$0.19 & 4,6      \nl 
1005$+$642 & GD 462       & 14.03 &         &         & 12       \nl 
1011$+$570 & GD 303       & 14.57 & $-$0.49 & $-$0.32 & 4,13     \nl 
1012$+$007 & PG           & 14.90 & $-$0.12 & $-$0.30 & 3,10     \nl 
\medskip                                                             
1019$+$637 & G 235$-$67   & 14.71 & $+$0.36 & $-$0.49 &          \tablebreak

1022$+$050 & LP 550$-$52  & 14.18 & $+$0.19 & $-$0.51 & 7        \nl 
1026$+$023 & LP 550$-$292 & 14.20 & $-$0.41 & $+$0.42 & 4        \nl 
1057$+$719 & PG           & 14.95 & \nodata & \nodata & 12       \nl 
1101$+$249 & Feige 36     & 12.70 & $-$0.23 & $-$0.89 & 3        \nl 
\medskip                                                             
1101$+$364 & Ton 1323     & 14.44 & $-$0.46 & $+$0.40 & 4,14     \nl 

1104$+$602 & G 197$-$4    & 13.80 & $-$0.02 & $-$0.81 & 1        \nl 
1121$+$216 & G 120$-$45   & 14.24 & $+$0.31 & $-$0.52 &          \nl 
1123$+$189 & PG           & 14.01 & $-$0.41 & $-$0.59 & 4,15     \nl 
1126$+$384 & GD 310       & 14.89 & $-$0.12 & $-$1.01 &          \nl 
\medskip                                                             
1129$+$156 & PG           & 14.04 & $-$0.52 & $+$0.11 & 4        \nl 

1204$+$451 & PG           & 14.84 & \nodata & \nodata & 12,20    \nl 
1210$+$533 & PG           & 14.07 & $-$0.70 & $-$0.73 & 4,5,22   \nl 
1229$-$013 & PG           & 14.24 & $-$0.35 & $+$0.05 & 4        \nl 
1232$+$479 & GD 148       & 14.52 & $+$0.06 & $-$0.68 & 1        \nl 
\medskip                                                             
1247$+$553 & GD 319       & 12.31 & $+$0.03 & $-$0.93 & 11,20    \nl 

1257$+$048 & GD 267       & 15.05 & $-$0.09 & $-$0.88 &          \nl  
1310$+$583 & PG           & 13.89 & \nodata & \nodata & 12       \nl 
1317$+$453 & G 177$-$31   & 14.13 & $+$0.03 & $-$0.56 & 1,16     \nl 
1319$+$466 & G 177$-$34   & 14.55 & $+$0.00 & $-$0.66 &          \nl 
\medskip                                                             
1408$+$323 & GD 163       & 14.06 & $+$0.03 & $-$0.81 & 1        \nl

1411$+$219 & PG           & 14.38 & $-$0.36 & $-$0.21 & 4,13?,15,20 \nl 
1422$+$095 & GD 165       & 14.32 & $+$0.14 & $-$0.59 & 7        \nl 
1432$+$159 & PG           & 13.90 & $-$0.12 & $+$0.12 & 3,10     \nl 
1446$+$286 & Ton 214      & 14.54 & $-$0.09 & $+$0.01 & 10       \nl 
\medskip                                                             
1459$+$305 & PG           & 13.98 & $+$0.30 & $-$0.84 & 20       \nl 

1537$+$652 & GD 348       & 14.64 & $+$0.18 & $-$0.49 &          \nl 
1538$+$269 & Ton 245      & 13.89 & $-$0.24 & $-$0.90 & 11       \nl
1559$+$369 & G 180$-$23   & 14.36 & $+$0.17 & $-$0.56 &          \nl 
1601$+$581 & PG           & 14.30 & \nodata & \nodata & 12       \nl 
\medskip                                                             
1615$-$154 & G 153$-$41   & 13.41 & $-$0.22 & $-$1.08 & 2        \nl 

1630$+$618 & GD 354       & 15.5  & CC$-$1  &         & 11       \nl
1632$+$177 & PG           & 13.08 & $-$0.13 & $+$0.47 & 4        \nl
1633$+$433 & G 180$-$63   & 14.82 & $+$0.42 & $-$0.42 &          \nl 
1637$+$335 & G 180$-$65   & 14.65 & $+$0.18 & $-$0.57 &          \nl 
\medskip                                                             
1647$+$591 & G 226$-$29   & 12.24 & $+$0.16 & $-$0.62 & 2        \nl 

1713$+$332 & GD 360       & 14.46 & $-$0.11 & $-$0.90 & 1,16     \nl 
1713$+$695 & G 258$-$3    & 13.27 & $+$0.07 & $-$0.70 & 1        \nl 
1716$+$426 & PG           & 13.85 & $-$0.66 & $-$0.35 & 3,4      \nl 
1756$+$827 & G 259$-$21   & 14.30 & $+$0.35 & $-$0.52 &          \nl 
\medskip                                                             
1822$+$410 & GD 378       & 14.39 & $-$0.19 & $-$0.96 & 1,13?,19?\tablebreak

1824$+$040 & G 21$-$15    & 13.90 & $+$0.05 & $-$0.55 & 1,7      \nl 
1826$-$045 & G 21$-$16    & 14.54 & $+$0.24 & $-$0.56 &          \nl
1840$+$042 & GD 215       & 14.73 & $+$0.25 & $-$0.62 &          \nl 
1845$+$019 & KPD          & 12.96 & $-$0.23 & $-$1.09 & 7        \nl 
\medskip                                                             
1919$+$145 & GD 219       & 13.01 & $+$0.06 & $-$0.66 & 1        \nl 

1935$+$276 & G 185$-$32   & 12.97 & $+$0.17 & $-$0.56 &          \nl 
1936$+$327 & GD 222       & 13.58 & $-$0.12 & $-$0.88 &          \nl 
1943$+$163 & G 142$-$50   & 14.08 & $-$0.06 & $-$0.86 &          \nl 
1953$-$011 & G 92$-$40    & 13.71 & $+$0.30 & $-$0.61 & 7        \nl 
\medskip                                                             
2009$+$622 & GD 543       & 15.15 & $-$0.69 & $-$0.29 & 4,17     \nl 

2028$+$390 & GD 391       & 13.37 & $-$0.15 & $-$0.98 &          \nl 
2032$+$188 & GD 231       & 15.34 & $-$0.04 & $-$0.80 & 16       \nl
2032$+$248 & Wolf 1346    & 11.54 & $-$0.07 & $-$0.87 & 1,2      \nl
2047$+$372 & G 210$-$36   & 12.93 & $+$0.14 & $-$0.66 &          \nl 
\medskip                                                             
2111$+$498 & GD 394       & 13.09 & $-$0.24 & $-$1.15 &          \nl 

2117$+$539 & G 231$-$40   & 12.33 & $+$0.07 & $-$0.67 & 2        \nl 
2124$+$550 & G 231$-$43   & 14.66 & $+$0.15 & $-$0.68 &          \nl 
2126$+$734 & G 261$-$43   & 12.88 & $+$0.01 & $-$0.66 & 2,18     \nl 
2134$+$218 & GD 234       & 14.45 & $-$0.04 & $-$0.83 &          \nl 
\medskip                                                             
2136$+$828 & G 261$-$45   & 13.02 & $-$0.02 & $-$0.72 & 1        \nl 

2149$+$021 & G 93$-$48    & 12.77 & $+$0.00 & $-$0.78 & 1,2,7    \nl 
2150$+$338 & GD 398       & 15.11 & $-$0.03 & $+$0.25 & 10       \nl 
2204$+$071 & PG           & 14.74 & $-$0.05 & $-$0.30 & 19       \nl 
2207$+$142 & G 18$-$34    & 15.61 & $+$0.32 & $-$0.58 &          \nl 
\medskip
2246$+$223 & G 67$-$23    & 14.31 & $+$0.21 & $-$0.67 &          \nl 

2307$+$636 & G 241$-$46   & 14.16 & $-$0.70 & $-$0.09 & 4        \nl 
2309$+$105 & GD 246       & 13.11 & $-$0.32 & $-$1.23 & 1,2,7    \nl
2319$+$691 & GD 559       & 14.75 & $-$0.07 & $+$0.23 & 10       \nl
2326$+$049 & G 29$-$38    & 13.10 & $+$0.20 & $-$0.65 & 2        \nl
\medskip
2328$+$108 & PG, KUV      & 15.53 & $-$0.03 & $-$0.90 &          \nl

2329$+$407 & G 171$-$2    & 13.82 & $+$0.03 & $-$0.72 & 1        \nl
2341$+$322 & G 130$-$5    & 12.92 & $+$0.14 & $-$0.61 & 1        \nl
2342$+$806 & GD 561       & 14.53 & $-$0.25 & $-$1.21 & 5,22     \nl
2345$+$318 & PG           & 14.18 & $-$0.05 & $+$0.01 & 3,10     \nl
\enddata
\tablerefs{
1) \cite{robshaft} 2) \cite{fossetal} 3) Saffer 1991; \cite{sbkl} 4)
Greenstein multichannel $v$, $g$--$r$, $u$--$v$ 5) \halpha\ emission
6) magnetic -- \cite{gdsps} 7) \cite{bragetal} 8) \cite{jhkkfw} 9)
\cite{trmduck} 10) Str\" omgren $y$, $b$--$y$, $u$--$b$ 11) sdB --
narrow He I \lam{4471} 12) PG $B$ 13) DBA 14) \cite{tmarsh1} 15) PG
COMP 16) \cite{tmarsh2} 17) \cite{szb} 18) \cite{zucketal} 19) DAB
20) Visual double 21) Photographic magnitude and color class --
\cite{giclas} 22) DAO
}
\end{deluxetable}                                                          

\begin{deluxetable}{clrrl}
\footnotesize
\tablewidth{300pt}
\tablecaption{Hot Subdwarf (sdB) Sample \label{sdbsample}}
\tablehead{
\colhead{WD}                & \colhead{Name}         & 
\colhead{$V_{\rm rad}$\,} & \colhead{$\sigma$~} & 
\colhead{Remark}
}
\startdata
0154$+$204 &            &  $+$13.0 & 30.9 & \nl
0154$+$182 &            &  $+$86.5 & 25.8 & \nl
0212$+$230 &            &  $-$78.2 & 15.4 & \nl
0250$+$189 &            &   $+$4.4 & 11.4 & \nl
\medskip		    		        
0322$+$114 &            &  $+$28.5 & 9.3  & \nl
		    
0342$+$026 &            &  $+$12.6 & 7.0  & \nl
0349$+$094 &            &  $+$30.4 & 10.8 & \nl
0749$+$658 &            &  $-$27.3 & 6.9  & \nl
0806$+$516 &            &   $+$5.6 & 14.4 & \nl
\medskip		    		        
0816$+$313 & Ton 313    &  $-$52.7 & 14.9 & \nl
		    
0823$+$465 &            &  $+$23.7 & 9.9  & \nl
0839$+$399 & K345$-$30  &   $-$9.3 & 13.2 & visual double \nl
0856$+$121 &            &  $+$97.0 & 10.2 & \nl
0918$+$029 &            &  $+$79.4 & 16.3 & \nl
\medskip		    		        
0919$+$272 & Ton 13     &  $-$33.2 & 16.9 & \nl
		    
1114$+$072 & Feige 38   &   $+$3.6 & 8.3  & \nl
1154$-$070 &            &   $-$4.7 & 13.2 & \nl
1223$+$058 &            & $+$131.4 & 21.5 & \nl
1224$+$671 &            &  $-$41.1 & 7.0  & \nl
\medskip		    		        
1230$+$052 &            &  $-$57.1 & 12.6 & \nl
		    
1233$+$426 & Feige 65   &  $+$61.2 & 13.4 & \nl
1303$-$114 &            &   $-$3.2 & 13.5 & \nl
1323$-$085 &            &  $-$53.2 & 7.8  & \nl
1325$+$101 &            &  $+$11.5 & 22.3 & \nl
\medskip		    		        
1343$-$101 &            &  $+$48.9 & 17.4 & \nl
		    
1442$+$342 &            &  $-$19.0 &  5.2 & \nl
1623$+$386 & KUV        &  $-$90.9 & 41.7 & visual double \nl
1643$+$209 &            &  $+$52.3 & 12.8 & \nl
1704$+$221 &            &  $-$62.1 & 9.2  & \nl
\medskip		    		        
1708$+$409 &            & $-$197.4 & 13.1 & \nl
		    
2059$+$013 &            &   $-$7.6 & 14.3 & \nl
2110$+$127 &            &   $+$6.8 & 5.2  & PG COMP \nl
2128$+$096 &            &  $+$23.0 & 19.4 & \nl
2135$+$044 &            &  $-$39.0 & 7.3  & \nl
\medskip		    		        
2204$+$034 &            &  $-$10.1 & 13.7 & \nl
		    
2214$+$183 &            &  $-$17.6 & 8.1  & \nl
2229$+$099 &            &  $-$19.3 & 9.9  & \nl
2301$+$259 &            & $-$130.5 & 9.4  & \nl
2356$+$166 &            &   $-$7.4 & 8.1  & \nl
\enddata
\end{deluxetable}                                                          

\begin{deluxetable}{cllcclc}
\footnotesize
\tablewidth{300pt}
\tablecaption{White Dwarf \& sdB Radial Velocity Variables
\label{ddvariables}} 
\tablehead{
\colhead{WD}         & \colhead{Name}     & \colhead{Reference}  &
\colhead{Weight}
}
\startdata
0101$+$039 & Feige 11     & 3       & 1    \nl 
0112$+$142 & PG           & 3,4     & 1    \nl 
0401$+$250 & G 8$-$8      & 1       & 2    \nl 
0549$+$158 & GD 71        & 1       & 2    \nl 
\medskip
0710$+$741 & GD 448       & 9       & 1    \nl 

0834$+$501 & PG           & 5       & 2    \nl 
0839$+$232 & PG           & 1       & 2    \nl 
0940$+$068 & PG           & 7,11,20 & 1    \nl 
1011$+$570 & GD 303       & 4,13    & 2    \nl 
\medskip
1012$+$007 & PG           & 3,10    & 2    \nl 

1019$+$637 & G 235$-$67   &         & 2    \nl 
1022$+$050 & LP 550$-$52  & 7       & 1    \nl 
1101$+$364 & Ton 1323     & 4,14    & 1    \nl 
1101$+$249 & Feige 36     & 3       & 1    \nl 
\medskip
1129$+$156 & PG           & 4       & 2    \nl 

1204$+$451 & PG           & 12,20   & 1    \nl 
1229$-$013 & PG           & 4       & 2    \nl 
1232$+$479 & GD 148       & 1       & 1    \nl 
1247$+$553 & GD 319       & 11,20   & 1    \nl 
\medskip
1310$+$583 & PG           & 12      & 1    \nl 

1317$+$453 & G 177$-$31   & 1,16    & 1    \nl 
1422$+$095 & GD 165       & 7       & 1    \nl 
1432$+$159 & PG           & 3,10    & 1    \nl 
1446$+$286 & Ton 214      & 10      & 2    \nl 
\medskip
1538$+$269 & Ton 245      & 11      & 1    \nl

1647$+$591 & G 226$-$29   & 2       & 1    \nl 
1713$+$332 & GD 360       & 1,16    & 1    \nl 
1716$+$426 & PG           & 3,4     & 1    \nl 
1824$+$040 & G 21$-$15    & 1,7     & 1    \nl 
\medskip
2009$+$622 & GD 543       & 4,17    & 1    \nl 

2117$+$539 & G 231$-$40   & 2       & 1    \nl 
2207$+$142 & G 18$-$34    &         & 1    \nl 
2345$+$318 & PG           & 3,10    & 1    \nl
\enddata
\end{deluxetable}                                                          

\begin{deluxetable}{clccccc}
\footnotesize
\tablewidth{350pt}
\tablecaption{Derived parameters of known close DDs. \label{knowndd}}
\tablehead{
\colhead{Object}    & \colhead{Name}                      & 
\colhead{$P$(d)}    & \colhead{log $P$(d)}  & 
\colhead{$M_1$}     & \colhead{$M_{\rm tot}$}             & 
\colhead{Reference}
}
\startdata                                                           
0101$+$039 & Feige 11  & 0.67 & $-$0.173 & 0.50 &      & 1       \nl 
           &           & 0.57 & $-$0.243 &      &      & alias?  \nl 
0135$-$052 & L 870--2  & 1.56 & $+$0.193 & 0.47 & 0.99 & 2       \nl 
0957$-$666 & L 101--26 & 0.06 & $-$1.234 & 0.37 & 0.69 & 3,4     \nl 
1101$+$249 & Feige 36  & 0.35 & $-$0.456 & 0.50 &      & 5       \nl 
1101$+$364 & Ton 1323  & 0.15 & $-$0.824 & 0.31 & 0.58 & 6       \nl 
1202$+$608 & Feige 55  & 1.49 & $+$0.173 & 0.40 &      & 7       \nl 
1241$-$010 &           & 3.35 & $+$0.525 & 0.31 &      & 8       \nl 
1317$+$453 &           & 4.80 & $+$0.681 & 0.33 &      & 8       \nl 
1432$+$159 & PG        & 0.23 & $-$0.649 & 0.50 &      & 1       \nl 
1538$+$269 & Ton 245   & 2.50 & $+$0.389 & 0.50 &      & 5       \nl 
1713$+$332 &           & 1.12 & $+$0.049 & 0.35 &      & 8       \nl 
2331$+$290 &           & 0.17 & $-$0.770 & 0.39 &      & 8       \nl 
2345$+$318 & PG        & 0.24 & $-$0.618 & 0.50 &      & 1       \nl 
\enddata
\tablerefs{
1) Saffer 1991; Saffer et al. 1994; Moran et al. 1998, in preparation
2) Saffer et al. 1988 3) \cite{brag95} 4) Moran et al. 1997 5) Saffer
et al. 1998, in preparation 6) Marsh 1995 7) Holberg et al. 1995 8)
Marsh et al. 1995
}
\end{deluxetable}

\begin{deluxetable}{cccc}
\footnotesize
\tablewidth{350pt}
\tablecaption{ Predicted Fractions of DDs and WD+MS Pairs
\label{ddtheory}} 
\tablehead{
\colhead{\ace}              & \colhead{Close DD/Total}     & 
\colhead{pre-\snia/Total}   & \colhead{Close WD+MS/Total}
}
\startdata
0.5 & 1/13    & 1/570 & 1/16 \nl
1.0 & 1/9\phn & 1/340 & 1/13 \nl
2.0 & 1/7\phn & 1/214 & 1/11 \nl
\enddata
\end{deluxetable}

\vfill\eject

\begin{deluxetable}{lcc}
\footnotesize
\tablewidth{250pt}
\tablecaption{ Detection Efficiencies (\%) \label{allddprob}}
\tablehead{
\colhead{}    & \colhead{~~$i = 90^\circ$}  & \colhead{$P(i) = \sin i$} 
}
\startdata
All $P$                            & 70 $\pm$ 4 & 64 $\pm$ 6 \nl
$P \in$ [2$^{\rm h}$,20$^{\rm d}$] & 79 $\pm$ 4 & 69 $\pm$ 5 \nl
$P \in$ [2$^{\rm h}$,10$^{\rm d}$] & 83 $\pm$ 3 & 74 $\pm$ 5 \nl
\enddata
\end{deluxetable}

\vfill\eject

\end{document}